\documentclass[prb]{revtex4-1}
\usepackage{dsfont}
\usepackage{amsmath}
\usepackage{enumerate}
\usepackage{amssymb}
\usepackage{graphicx}
\usepackage{wrapfig}
\usepackage{setspace}

\def\XXint#1#2#3{{\setbox0=\hbox{$#1{#2#3}{\int}$}
     \vcenter{\hbox{$#2#3$}}\kern-.5\wd0}}

\renewcommand{\>}{\rangle}
\newcommand{\<}{\langle}

\begin{document}

\title{Exact expectation values within Richardson's approach for the pairing Hamiltonian in a macroscopic system}
\author{G. Gorohovsky}
\affiliation{Racah Institute of Physics, Hebrew University,
Jeruslaem, Israel}
\author{E. Bettelheim}
\affiliation{Racah Institute of Physics, Hebrew University,
Jeruslaem, Israel}

\date{\today}

\begin{abstract}
BCS superconductivity is explained by a simple Hamiltonian describing an attractive pairing interaction between pairs of electrons. The Hamiltonian may be treated using a mean field method, which is adequate to study equilibrium properties and a variety of non-equilibrium effects. Nevertheless, in certain non-equilibrium situations, even in a macroscopic, rather than a microscopic, superconductor, the application of mean field may not be valid. In such cases, one may resort to the full solution of the Hamiltonian, as given by Richardson in the 60's. The relevance of Richardson's solution to macroscopic non-equilibrium superconductors was pointed out recently based on the existence of quantum instabilities out of equilibrium. It is then of interest to obtain analytical expressions for expectation values between exact eigenvalues of the pairing Hamiltonian within the Richardson approach for macroscopic systems. We undertake this task in the current paper. It should be noted that Richardson's approach yields the full set of eigenvalues of the Hamiltonian, while BCS theory yields only a subset. The results obtained here, then, generalize the familiar BCS expressions for, e.g., the electron occupation or pairing correlations, to cases where the spectrum of excitations diverges from BCS theory, for example in cases where the spectrum exhibits multiple gaps.
\end{abstract}

\maketitle

\section{Introduction and Results}

Our basic understanding of superconductivity is informed by the mean field solution of the pairing Hamiltonian, Eq. (\ref{Hamiltonian}), given by Bardeen-Cooper-Schrieffer\cite{BCS1, BCS2} (BCS) some half a century ago. The mean field solution does extremely well in many respects because of the fact that, in essence, the condensate interacts equally strong with all energy levels that  participate in superconductivity. Thus a macroscopic system is well described by mean field theory. There are some caveats, though. The BCS expression for the eigen-states does not capture all possible eigen-states of the pairing Hamiltonian. For example the BCS expression predicts a single gap in the excitation spectrum, while as it turns out from an exact solution, any number of gaps may appear in the spectrum. This is known because Richardson\cite{Richardson} has solved the pairing Hamiltonian exactly. These unexpected eigen-states, which correspond to an unusual spectra of excitations do not play an important role in equilibrium -- a stroke of luck for BCS theory. Nevertheless they may appear in out of equilibrium situations, when the system is far enough from equilibrium\cite{Bettelheim:Multi:Gapped}, this is due to certain quantum instabilities encountered in non-equilibrium superconductivity  \cite{spivak}.

In this paper we study more closely the consequences of these multi-gapped eigen-states. We make use of Richardson's exact solution and Slavnov's formula\cite{Slavnov}  as applied to this model \cite{Zhou:Links:McKenzie:Gould} in order to compute  correlation functions of the pairing amplitude and the occupation number at different energy levels for a macroscopic system analytically. It should be noted that Richardson already derived expressions for correlation functions, which are specific cases of Slavnov's formulas \cite{Richardson:Unrestricted} deriving their thermodynamic limit for the case of one gap and for the expectation value of a single operator\cite{Richardson:Large:N}. Our a result is a generalization of that work, giving expressions for any number of operators both in the cases of one {\it or two} spectral gaps. Similar approaches were used in [\onlinecite{Calabrese, Amico:Osterloh:Exact:Correlations}] using numerics and focusing rather on mesoscopic systems. The relevance of Richardson's exact solution to the pairing problem in mesoscopic systems was pioneered in Ref. [\onlinecite{Braun:VonDelft}]. Our results, dealing with a macroscopic system,  agree with  BCS theory, when there is only one spectral gap. Indeed, the expectation value in that case is given by Eq. (\ref{BCSResult}), below, (where $R_2(\xi) = \sqrt{(\xi-\mu)^2 + \Delta^2}$), an expression which is familiar from BCS theory.

\section{The Richardson Solution}
BCS superconductivity is captured by a model Hamiltonian, designed to include only those features, that are crucial to the existence of superconductivity. The Hamiltonian includes a free, bilinear, part composed of $L$ single particle levels and an interaction part which scatters Cooper pairs:
\begin{align} \label{Hamiltonian}
H = \sum_{\begin{array}{c} 1\leq j \leq L \\ \sigma_j=\pm \end{array}} \varepsilon_j c^\dagger_{j,\sigma_j} c_{j,\sigma_j} - G \sum_{\begin{array}{c} 1\leq j \leq L \\ 1\leq l \leq L \end{array}} c^\dagger_{j,+} c^\dagger_{j,-} c_{l,+}c_{l,-}.
\end{align}
Here $(j,+)$ and $(j,-)$ denote the quantum numbers of time reversed pairs. For example, if $(j,+)$ denotes a state with wave number $\vec{k}$ and spin up, then $(j,-)$ denotes a state with wave number $-\vec{k}$ and spin up. We assume, for simplicity, that each level $j$ is only doubly degenerate, where $\sigma$ indexes the two degenerate states, $\sigma$ taking $+$ and $-$ as values. Furthermore, we assume uniform level spacing $\varepsilon_j  - \varepsilon_{j-1} = \delta$.

It turns out that the model Hamiltonian is exactly solvable, namely, the eigenvalues and eigenstates can be found exactly. This was done by Richardson in the 60's \cite{Richardson}, motivated by the Hamiltonian's importance in nuclear physics. The solution of the Hamiltonian is not trivial, and in fact the Richardson solution may be understood as a Bethe ansatz solution. Indeed, the Hamiltonian contains a non-trivial interaction which scatters Cooper pairs (namely time reverse pairs of electrons). Note however, that those electrons which  singly occupy levels do not scatter. The levels which contain singly occupied states are then called {\it blocked levels}, since no pair can scatter into them. The set of levels which are occupied by single electrons, together with the corresponding spins of the electrons, are good quantum numbers. A given set of such quantum number is termed {\it seniority}. For each given seniority, denoted by $(j_1, \sigma_1, j_2, \sigma_2, \dots, j_M, \sigma_M)$, one may define a 'vacuum' state $|\{j_i,\sigma_i\}_{i=1}^M \>$, which contains no pairs, but only singly occupied levels:
\begin{align}
|\{j_i,\sigma_i\}_{i=1}^M \>  = \prod_i c^\dagger_{j_i,\sigma_i} |0\>
\end{align}

Another good quantum number is the number of rapidities, $P$. The total number of electrons in the state is then $2P+M$, since each rapidity contributes one pair of electrons, due to Eq. (\ref{RichardsonState}). There is a one to one correspondence of eigenstates with given $P$ and given seniority and solutions, $\{E_\nu\}_{\nu=1}^P$, of the Richardson equations:
\begin{align}\label{RichardsonEq}
 \sum_{\mu\neq \nu} \frac{1}{ E_\nu-E_\mu}  - \frac{1}{2} \sum_{ i \in U } \frac{1}{E_\nu - \varepsilon_j }  - \frac{1}{G} =0 ,
\end{align}
Here $U$ is the set of all unblocked levels, namely $U = \{i | \nexists k, i=j_k\}$.
The eigenstate is then denoted by $| \{E_i\}_{i=1}^P , \{j_i , \sigma_i \}_{i=1}^M  \>$ and is given explicitly by:
\begin{align} \label{RichardsonState}
| \{E_\nu\}_{\nu=1}^P , \{j_i , \sigma_i \}_{i=1}^M  \> = \prod_{\alpha=1}^{P} b^\dagger_\alpha |\{j_i,\sigma_i\}_{i=1}^M \>,\quad b^\dagger_\alpha = \sum_{ i \in U} \frac{1}{E_\alpha - \varepsilon_i} c^\dagger_{i, \uparrow} c^\dagger_{i, \downarrow}
\end{align}
The eigenvalue for the state is $E = \sum_\nu 2 E_\nu + \sum_{j\notin U} \varepsilon_j$.

We shall term the $E_\nu$'s as {\it rapidities} characterizing the state $| \{E_\nu\}_{\nu=1}^P , \{j_i , \sigma_i \}_{i=1}^M  \>$. We shall term a state of the form (\ref{RichardsonState}) a 'Richardson state' even if the rapidities do not satisfy (\ref{RichardsonEq}). Namely, a Richardson state is an eigenstate if and only if the rapidities satisfy (\ref{RichardsonEq}).

\section{Electrostatic Analogy}
In what follows we shall be interested in computing expectation values in the Richardson model in the continuum limit. To do so, Richardson's equations (\ref{RichardsonEq}) must be solved in different circumstances. The solution of the Richardson equations are facilitated by the fact that these have a convenient 2D electrostatic interpretation. The electric field at point $w$  of a charge placed in at point $z$ in the complex plane is  given by $E_x - i E_y = \frac{1}{w-x}$, which means that (\ref{RichardsonEq}) may be interpreted as the condition of electrostatic equilibrium of the charges $E_\nu$, which are assigned a charge $+1$, given the the position of charges of magnitude $-1/2$ at $\varepsilon_i$ (only for $i\in U$ ) and a constant field, $\frac{1}{g}$, pointing in the negative $x$ direction.

Given this electrostatic analogy it is natural also to define the field $h(z) = \delta\left( E_x(z) + i E_y(z) \right)$ (here $\delta$ is the level spacing). Explicitly, $h(z)$ is given by:
\begin{align} \label{hDefinition}
h(\xi) = - \frac{\delta}{2} \sum_{j\in U} \frac{1}{\xi - \varepsilon_j} + \delta \sum_\nu \frac{1}{\xi-E_\nu} - \frac{1}{g}.
\end{align}
Here $g=\frac{G}{\delta}$. The continuum limit is taken by letting $\delta \longrightarrow 0$ while letting $h(z)$ tend to a constant, except for certain arcs and a segment of the real axis. In fact the solutions of Richardson's equations have the following form in the continuum limit: as can be easily understood, there is always an electrostatic equilibrium position between any two adjacent unblocked $\varepsilon_j$'s. As the level spacing is decreased $\delta \to 0$, one may create any line density of charges on the real axis, by placing rapidities on the real axis between adjacent unblocked levels or by blocking levels. This defines a coarse grained charge, $\rho(\varepsilon)$ on the real axis, given by
\begin{align}
\rho(\varepsilon) = \delta \cdot \left[ \sum_{\nu, E_\nu \in \mathbb{R}} \tilde{\delta}_\delta(\varepsilon-E_\nu) - \frac{1}{2} \sum_{i, \nexists k ,j_k= i} \tilde{\delta}_\delta(\varepsilon-\varepsilon_i) \right],
\end{align}
here $\tilde{\delta}_\delta$ is a function tending to a delta-function as $\delta \to 0$ having a width much larger than $\delta$, e.g. $\tilde{\delta}_\delta(x) = \frac{1}{\sqrt{\pi} A \delta} e^{\left( \frac{x}{A \delta} \right)^2}$, for some large number $A$. In addition to those rapidities, which are on the real axis, and lie between the $\varepsilon$, there may exist truly complex rapidities. These, it turns out, arrange themselves on arcs. The arcs extend symmetrically around the real axis (because the Richardson equations are real, complex rapidities come in complex conjugate pairs). Suppose there are $K$ arcs. We shall denote the two endpoints of arc $j$ as $\mu_j \pm i \Delta_j$ (here $i$ is the imaginary unit).

The field $h(\xi)$ consequently has a jump discontinuity on the real axis of magnitude $\rho(\varepsilon)$ and on the arcs, where it has some $O(1)$ jump discontinuity, which must be determined.  Consider the endpoints of the arcs. Those rapidities that sit on the endpoints, must be in electrostatic equilibrium. A closer analysis shows that this is only possible if the field $h(\xi)$ as $\xi$ approaches the endpoint tends to $0$. This is an intuitive result, since if $h(\xi)$ would not approach $0$ the endpoints would feel a force that would move them. One concludes that $h(\xi)$ vanishes on the 2K endpoints of the arcs. Moreover, if we look at the average value of $h(\xi)$ across the arc (namely $\frac{h(\xi_+)+h(\xi_-)}{2}$, where $\xi_\pm$ are points just to the left and to the right of the arc respectively), then this average must vanish. The reason being, that this average represents the far-field felt by the charges on the arc. Looking under a magnifying glass at a segment of the arc, one sees a long (from this perspective, infinite) chain of charges. These chains will fly off if an external (or far-) field is present. Namely, the average must vanish. These considerations allow finding $h(\xi)$. In fact Gaudin in a paper in French\cite{Gaudin} (later reviewed and expanded in English in [\onlinecite{Sierra}]) had shown that it is given by
\begin{align} \label{hIntegral}
h(\xi) = R_{2K}(\xi) \int \frac{ \rho(\varepsilon)}{R_{2K}(\varepsilon)(\varepsilon-\xi)} d\varepsilon,
\end{align}
where
\begin{align}\label{Rdefinition}
R_{2K}(\xi) = \sqrt{\prod_{j=1}^K (\xi - \mu_j)^2 +\Delta_j^2}
\end{align}
Indeed, $h(\xi)$ defined by (\ref{hIntegral})  has a jump continuity on the real axis of the given value $\rho(\varepsilon)$, vanishes on the endpoints of the arc has a non-trivial jump discontinuity on  $K$ arcs, and  its average value across an arc is $0$ (since it simply changes sign across the arc).

Eq. (\ref{hIntegral}) is an expression for $h(\xi)$ given a knowledge of $\rho(\varepsilon)$ and of the endpoints of the arcs $\mu_j, \Delta_j$, $j=1,\dots,K$. $\rho(\varepsilon)$ is arbitrary and may be tuned by  blocking levels and placing rapidities between adjacent unblocked levels. The arc endpoints must be determined self-consistently, however. These self-consistency conditions can be derived by noting that $h(\xi)$ as defined by (\ref{hDefinition})  must have the following asymptotic behavior as $\xi\to\infty$: $h(\xi) \to \frac{1}{g} +O\left(\frac{1}{\xi} \right)$. Expanding in large  $\xi$, Eq. (\ref{hIntegral}) shows that the expected asymptotic behavior of $h(\xi)$ is only satisfied if the following $K-1$ conditions hold:
\begin{align}\label{selfconsistencies}
\int \frac{\rho(\varepsilon) \varepsilon^l }{R_{2K}(\varepsilon)} d\varepsilon = \frac{1}{g}\delta_{l,K-1} , \quad l\leq K-1.
\end{align}
These are not enough to determine $2K$ free parameters which determine the position of the endpoints. Extra conditions may be found if one knows the number of rapidities on each one of the arcs. In the case of one arc, the number of rapidities on the arc is known if one knows the total number of electrons. Indeed, the total number of particles is $2P+M$, where $P$ is the number of rapidities and $M$ are the number of singly occupying electrons. The number of rapidities on the real axis is known since we know $\rho(\varepsilon)$ and so the number of rapidities on the arc is also known. If there is more than one arc, however, the total number of particles is not enough to fully determine the endpoints, and for the same number of particles, the same given $\rho(\varepsilon)$ and the same number of arcs, one may find different solutions depending on how many rapidities occupy each arc. The solutions differ by the location of the endpoints of the arcs.

Since the total number of particles is a good quantum number, which also factors into finding the endpoints of the arc, it is useful to have an expression for this quantity. In fact we shall want to compute $J_z  = \delta \frac{2P + M - L}{2}$. This quantity is directly related to the total number of particles, $J_z = \delta \left\<\sum_i \left( \frac{\hat{N}_i-1}{2}\right)\right\>$. $J_z$ in fact features in the asymptotics of $h(\xi)$ as $\xi \to \infty$.  Indeed, as $\xi \to\infty$, $h(\xi)  \to \frac{1}{g} + \frac{J_z}{\xi}$. Expanding Eq. (\ref{hIntegral}) for large $\xi$, one obtains:
\begin{align}\label{JzAsIntegra}
 J_z=   \int  \frac{\left[ R_{2K}(\xi) \right]_+  \rho(\xi) }{R_{2K}(\xi)} d\xi
\end{align}
where $\left[ f(\xi) \right]_+$ denotes the positive Laurent series of $f(\xi)$ when expanded around infinity.

The total energy may also be found. Instead of the total energy, E, we compute a slightly different but directly related quantity, $\mathcal{E}$, defined as follows:
\begin{align}\label{mathcalEdefinition}
\mathcal{E} = \frac{\delta}{2}\left(  E - \sum_{i=1}^L \varepsilon_i \right),
\end{align}
then $h(\xi) \to \frac{1}{g} + \frac{J_z}{\xi} + \frac{\mathcal{E}}{\xi^2}$

\begin{align} \label{mathcalEIntegral}
\mathcal{E}=   \int  \frac{\left[ \xi R_{2K}(\xi)  \right]_+  \rho(\xi) }{R_{2K}(\xi)} d\xi
\end{align}

The eigen-states described by Eq. (\ref{RichardsonState}) generalize the eigen-states found by BCS. States directly corresponding to the BCS eigen-states may be recovered, but in addition to the states, one finds states which have no BCS counterpart. To obtain the BCS eigen-states one must assume $K=1$. The expression for the total number of particles, the energy and the constraint, Eqs. (\ref{JzAsIntegra}), (\ref{mathcalEIntegral}) and (\ref{selfconsistencies}), respectively specialize to:
\begin{align}
J_z = \int \frac{(\varepsilon - \mu)\rho(\varepsilon)}{\sqrt{(\varepsilon-\mu)^2+\Delta^2}} d\varepsilon, \quad \mathcal{E} = \int \varepsilon \frac{(\varepsilon - \mu)\rho(\varepsilon)}{\sqrt{(\varepsilon-\mu)^2+\Delta^2}} d\varepsilon + \frac{\Delta^2}{2g}, \quad \frac{1}{g} = \int \frac{\rho(\varepsilon)}{\sqrt{(\varepsilon-\mu)^2+\Delta^2}} d\varepsilon.
\end{align}
These expressions are the familiar BCS expressions, if $\rho(\varepsilon)$ is identified as $\frac{n(\varepsilon)-1}{2}$, where $n(\varepsilon)$ is the occupation number of excitations at energy $\varepsilon$. Indeed, blocking a level or inserting an $E$ on the real axis change the energy of the system, and thus may be viewed as excitations. $\Delta$ is the size of the spectral gap and $\mu$ is the chemical potential of the condensate, to be determined self consistently, given the total number of particles in the system. It is not clear however that the Richardson wave function, Eq. (\ref{RichardsonState}), becomes the BCS state (or the number projected version thereof) in the thermodynamic limit. In particular, the factorized form of the BCS state does not appear naturally in Eq. (\ref{RichardsonState}). We shall see, however, that when $K=1$ all correlation functions factorize, thus confirming BCS results. We shall also derive the results for $K=2$, where, we show, the factorizability of correlation functions no longer holds.

\section{Spherical and Elliptical cases \label{SpherAndEllipt}}
In the above, we have shown how  to obtain the continuum solution of the Richardson equations and how to relate the solution  to the good quantum numbers $J_z$ and $\mathcal{E}$. This was based on the electrostatic solution given by Gaudin\cite{Gaudin}. We are interested in finding expectation values.  The simplest expectation value to find is that of $\< \hat{S}^z_i \>$. Where
\begin{align}\label{SzDefinition}
\hat{S}^z_i =  c^\dagger_{i,+} c_{i,+}+c^\dagger_{i,-} c_{i,-}-1
\end{align}
By Hellman-Feynman, $\<\hat{S}^z_i\> =\frac{2}{\delta} \frac{\partial \mathcal{E}}{\partial \varepsilon_i}$, where $\mathcal{E}$ is given by (\ref{mathcalEIntegral}). More complicated expectation values are not simply computable by invoking Hellman-Feynman, still objects similar to $\frac{\partial \mathcal{E}}{\partial \varepsilon_i}$ will appear in such computations. More precisely, we have
$ \frac{\partial \mathcal{E}}{\partial \varepsilon_i} = \delta \left( -\frac{1}{2} + \sum_\nu \frac{\partial E_\nu}{\partial \varepsilon_i} \right)$,  and the object that will repeatedly appear in subsequent calculations will  be $\frac{\partial E_\nu }{\partial \varepsilon_j}$. We turn, then, to the computation of this object.

It is easy to obtain information on $\frac{\partial E_\nu }{\partial \varepsilon_j}$, by considering the potential function $\phi(\xi)$, given by:
\begin{align}
\phi(\xi) = -\frac{\delta}{2} \sum_j \log(\xi - \varepsilon_j)  + \delta \sum_\nu \log( \xi - E_\nu)
\end{align}
Then, a coarse-grained $\frac{\partial E_\nu}{\partial \varepsilon_i}$  times the density of $E$'s is given by the jump discontinuity of $\frac{\partial \phi(\xi)}{\partial \varepsilon_i}$ at $\xi = E_\nu$, divided by $2\pi i \delta$. We compute then the potential in two cases: the case of one arc, and the case of two arcs.

Note that the solution (\ref{hIntegral}) derives its algebraic properties from the function $R_{2K}(\xi)$ defined in (\ref{Rdefinition}). $R_{2K}(\xi)$ in fact defines a spherical double-sheeted algebraic Riemann surface in the one arc case an an elliptical surface in the two arc case.  We shall use basic notions of algebraic geometry, especially in the two-arc or elliptical case, where direct manipulation becomes cumbersome, and theorems of uniqueness of functions on Riemann surfaces become a much more powerful route to obtain the results.

We are interested in finding the change in electrostatic potential $\delta \phi(\xi)$ when one changes the charge density on the real axis. It will be easier to compute first $\delta h(\xi) d\xi$ and then integrate.  We first compute $\delta h(\xi) d\xi$ when a unit charge is added at point $\varepsilon$ on the real axis, and then obtain a general result through a simple application of the superposition principle. Based on its definition, Eq. (\ref{hDefinition}), the field $h(\xi) d\xi$ will have a pole at $\varepsilon$, with residue $1$. Similarly, there will be a pole at infinity. Indeed, $\oint h(\xi) d\xi = \delta( N -\frac{L}{2})$, where the integration contour encircles infinity. In addition, the arc will move, namely $\Delta_i$ and $\mu_i$ will change. If we look at Eq. (\ref{hIntegral}) we see that, no matter how $\Delta_i$ and $\mu_i$ change, $h(\xi)$ will retain the properties that it is an algebraic function defined on the Riemann surface of $R_{2K}(x)$ and that it changes sign going from the upper sheet to the lower sheet. This means that $\delta h(\xi)d\xi$ will also have a pole at the lower sheet at $\varepsilon$ and at $\infty$  but with residues having reversed signs, respectively. We have $\delta h(\xi) d\xi$ as a meromorphic differential with given pole structure. Such a differential is unique on a spherical Riemann surface, and in fact is given by:
\begin{align}
\delta h (\xi) d\xi =  \frac{1}{\sqrt{(\xi - \mu)^2 +\Delta^2}}\left(1+ \frac{\sqrt{(\varepsilon - \mu)^2 +\Delta^2}}{\xi - \varepsilon} \right) d\xi
\end{align}
To obtain then $\delta h(\xi)$ for a generic disturbance of the charge density on the real axis, one must employ the superposition principle:
\begin{align}\label{deltaHSpherical}
\delta h = \int  \frac{\delta\rho(\varepsilon) }{\sqrt{(\xi - \mu)^2 +\Delta^2}}\left(1+ \frac{\sqrt{(\varepsilon - \mu)^2 +\Delta^2}}{\xi - \varepsilon} \right) d \varepsilon
\end{align}

Our goal is to obtain $\frac{\partial \phi(\xi)}{\partial \varepsilon_i}$. To obtain this we must assume that $\delta \phi(\xi)$ in (\ref{deltaHSpherical}) corresponds to an infinitesimal change produced by moving $\varepsilon_i$. It is important to realize, though, that when one moves a single $\varepsilon_i$ the rapidities present near it on the real axis will also move, thus contributing to $\delta \rho$. The computation of $\delta \rho$ when one moves a single $\varepsilon_i$ is then very difficult, since one has to know the exact configuration of rapidities near $\varepsilon_i$. This is not necessary, however, since we will be interested in coarse grained quantities. Indeed consider changing not a single $\varepsilon_i$ but rather a number, $A$,  of them centered around $\varepsilon$. It is easy to see that to leading order in $\delta$, as we shift this group of $\varepsilon_i$'s, the whole charge density, including the rapidities 'trapped' between the $\varepsilon_i$'s will shift rigidly. This implies the following equation for the change of charge occupation $\delta \rho$:
\begin{align}\label{deltaRho}
\delta \rho(\varepsilon') =  \delta \rho(\varepsilon) \delta'(\varepsilon' - \varepsilon) \delta \varepsilon \times \delta,
\end{align}
where $\frac{\delta \varepsilon}{A}$ is the amount each one of the $A$ $\varepsilon_i$'s around $\varepsilon$ was shifted. Plugging this into (\ref{deltaHSpherical}) we obtain:
\begin{align}
\frac{\delta h(\xi)}{\delta \varepsilon} = \delta \cdot \rho(\varepsilon) \frac{\Delta^2 +( \varepsilon-\mu ) (\xi - \mu) }{ {R_2(\varepsilon) R_2(\xi)} ( \varepsilon - \xi )^2 },
\end{align}
which upon integration yields:
\begin{align}\label{dphidepsiOneArc}
\frac{\delta \phi(\xi) } {\delta \varepsilon} = \delta \frac{\rho(\varepsilon )}{\varepsilon - \xi }  \frac{R_2(\xi)}{R_2(\varepsilon)}.
\end{align}

We wish now to obtain a similar result when two arcs are present. This requires working with the elliptic Riemann surface described by $R_4(\xi)$. Being elliptic, this Riemann surface has the topology of a torus. We may use  a rectangle with points on opposite sides identified (cyclic boundary conditions)  as a model for the torus. In other words, we shall write the results first on the rectangle and then map them to the algebraic curve defined by $R_4(\xi)$. Consider then a rectangle of sides $\omega_1 \in \mathbb{R}^+$ and $- i \omega_2 \in \mathbb{R}^+$. A function that maps this torus into the two sheeted algebraic curve defined by $R_4(\xi)$, is given by the inverse Abel map, written in terms of Weierstrass's elliptic Zeta function:
\begin{align}\label{xiofu}
\xi (u) = \zeta\left( u - u_\infty \right. \left| \omega_1, \omega_2 \right) - \zeta\left( u + u_\infty \right. \left| \omega_1, \omega_2 \right) + c,
\end{align}
while the direct map is given by:
\begin{align}
u(\xi) = \int^\xi \frac{d\xi'}{R_4(\xi')} .
\end{align}
This procedure is standard in the study of algebraic Riemann surface, being reviewed in any number of textbooks.

Consider now $\delta h(\xi) d\xi$ for the two-arc case. Being completely general, the pole structure is the same as in the one-arc case, and just as before starting from (\ref{hIntegral}) we may conclude that $\delta h(\xi) d\xi$ is an elliptic differential that changes sign as one changes sheets. We need one more ingredient to find $\delta h(\xi) d\xi$, as these conditions are not sufficient to determine a differential on an {\it elliptic} Riemann surface, due to the existence of the holomorphic differential, $du(\xi)$. An extra condition on is obtained by considering that the change performed must leave the number of pairs on each arc constant. Now, the number of pairs on an arc is proportional to $\oint \delta h(\xi) d\xi$, where the integral is taken around the arc, which serves as the required additional condition. The image of the arc on the rectangle is a line extending from $0$ to $\omega_1$. So we must demand that the integral of $\delta h(\xi) d\xi$ be zero taken on a line from $0$ to $\omega_1$. The final result is then:
\begin{align}\label{deltahElliptical}
 &\delta h(\xi) d \xi =  \\
  & = du(\xi) \int  \left( \zeta\left( u(\xi) - u(\varepsilon)\right) -  \zeta\left( u(\xi) - u_\infty \right)  - \zeta\left( u(\xi) + u(\varepsilon)  \right) + \zeta\left( u(\xi) +  u_\infty \right)  + \frac{4 (u(\varepsilon)-u_\infty) \zeta\left(\frac{\omega_1}{2}\right)}{\omega_1} \right) \delta \rho (\varepsilon) d \varepsilon \nonumber
\end{align}
Indeed, it is easy to verify that the expression inside the brackets in the integrand has poles at $u(\varepsilon)$ and $u_\infty$, which are just the images of $\varepsilon$ and $\infty$, respectively; The residues are correct; The integrand is invariant upon switching sheets (this is effected by taking $u(\xi) \to - u(\xi)$) but the integral is multiplied by $du$, which does change signs when one switches sheets, so that the whole expression has the right behavior upon switching sheets; It is also easy to verify, that the integral of this expression taken from $0$ to $\omega_1$ gives zero.

Plugging (\ref{deltaRho}) into (\ref{deltaHSpherical}) amounts to taking a derivative with respect to $u(\varepsilon)$, which yields:
\begin{align}
\delta h(\xi) d\xi = \rho(\varepsilon) \left [ \wp (u(\xi) - u(\varepsilon ) ) + \wp ( u(\xi) + u(\varepsilon ) ) + 4 \omega_1^{-1} \zeta \left(\frac{\omega_1}{2} \right)  \right] du(\xi) \delta u(\varepsilon )
\end{align}
It is a matter of performing an integral w.r.t $u(\xi)$ in order to obtain $\frac{\delta \phi}{\delta \varepsilon}$:
\begin{align}
\frac{\delta \phi(\xi)}{\delta \varepsilon} \delta \varepsilon = \rho(\varepsilon) \left [ \zeta (u(\xi) - u(\varepsilon ) ) + \zeta ( u(\xi) + u(\varepsilon ) ) + 4 \omega_1^{-1} u(\xi) \zeta \left(\frac{\omega_1}{2} \right)  \right]  \delta u(\varepsilon )
\end{align}
Using the identity:
\begin{align}
\left[ \zeta (u(\xi) - u(\varepsilon ) ) + \zeta ( u(\xi) + u(\varepsilon ) ) - \zeta (u(\xi) - u_\infty)  - \zeta ( u(\xi) + u_\infty )  \right] du(\varepsilon)= \frac{R_4(\xi) d\varepsilon}{R_4(\varepsilon)(\xi - \varepsilon)}
\end{align}
The expression for $\frac{\delta \phi}{\delta \varepsilon}$ may be written in a form which will prove much more convenient later:
\begin{align}\label{dphidepsilon2Gap}
\frac{\delta \phi (\xi)}{\delta \varepsilon }  = \delta \cdot \frac{\rho(\varepsilon )R_4(\xi)}{R_4(\varepsilon )}  \left( \frac{1}{\xi-\varepsilon } - g(\xi) \right),
\end{align}
where
\begin{align}\label{gDefinition}
g(\xi) = \frac{ \zeta(u(\xi) - u_\infty) +  \zeta(u(\xi)+u_\infty) - 4 \omega_1^{-1} u(\xi) \zeta\left(\frac{\omega_1}{2} \right)}{R_4(\xi)}.
\end{align}

\section{Slavnov's Formula}
In order to compute the matrix elements we shall make use of Slavnov's formula\cite{Slavnov}. Two states will have a non-zero overlap only if they have the same seniority, namely the same set of  singly occupied levels $j_i$ with the spins pointing in the same direction. We thus suppress the notation of seniority and write simply $|\{E_\nu\}_{\nu=1}^P \>$ for a Richardson state. Slavnov's formula\cite{Slavnov} as applied\cite{Zhou:Links:McKenzie:Gould} to the Richardson solution reads:
\begin{align}\label{Slavnov}
\< \{w_\nu\}_{\nu=1}^P | \{ v_\nu \}_{\nu=1}^P \>  = \frac{\prod_{a\neq b} (v_b - w_a )}{\prod_{b<a} (w_b-w_a) \prod_{a<b}(v_b-v_a)} \det J,
\end{align}
where $\{v_\nu\}_{\nu=1}^P$  obey the Richardson equations, while $\{w_\nu\}_{\nu=1}^P$   do not necessarily satisfy the Richardson equations. The matrix $J$ appearing in (\ref{Slavnov}) is given by:
\begin{align}\label{JDefinition}
J_{ab} = \frac{v_b-w_b}{v_a-w_b} \left(\sum_{\alpha=1}^P \frac{1}{(v_a-\varepsilon_\alpha)(w_b-\varepsilon_\alpha)} -2 \sum_{c\neq a} \frac{1}{(v_a-v_c)(v_b-v_c)} \right)
\end{align}
When the set $\{v_\nu\}_{\nu=1}^P$ coincides with the set $\{w_\nu\}_{\nu=1}^P$, Slavnov's formula gives the norm of the Richardson state. In this case the matrix $J$ takes the form:
\begin{align}\label{ADefinition}
A_{ab} = \left\{ \begin{array}{lr}
\sum_\alpha \frac{1}{(v_a-\varepsilon_\alpha)^2} - 2 \sum_{c\neq a}\frac{1}{(v_a-v_c)^2}  & a=b \\
\frac{2}{(v_a-v_b)^2}  & a\neq b
\end{array} \right. .
\end{align}
This limit form, $A$, of $J$ will appear frequently in the sequel.
\section{Computation of Expectation values: Basic examples \label{computeExpectation}}
The notations for the computation of a general correlation function become quite cumbersome. It is easier to start with two simple examples which demonstrate the principle of the computation before plunging into the general scheme. This is undertaken in the next two subsections, respectively.

\subsection{Computation of $\<\hat{S}_z(\varepsilon)\>$ \label{ComputeSz}}
Consider the computation of $\< \hat{S}_z(\varepsilon) \>$. $\hat{S}^z_\alpha$ is given by  (\ref{SzDefinition}). What is meant by $\hat{S}_z(\varepsilon)$ is a coarse grained version of this quantity, namely
\begin{align}\label{SzCoarseGrained}
\hat{S}_z(\varepsilon) = \frac{1}{2A} \sum_{|\varepsilon_i - \varepsilon|<A \delta} \hat{S}^z_i.
\end{align}
To compute such an object we may first compute simply $\< \hat{N}_i \>$ and then perform a coarse graining and subtract a constant to obtain $\< \hat{S}_z (\varepsilon) \>$.

We shall want to represent $\< \hat{N}_i \>$ as an overlap between two Richardson states, in order to use Slavnov's formula to compute it. More explicitly, we are computing $\< \{v_\nu\}_{\nu=1}^P | \hat{N}_i| \{v_\nu\}_{\nu=1}^P \>$, where $\hat{N}_i = \frac{c^\dagger_{i,+} c_{i,+} + c^\dagger_{i,-} c_{i,-}}{2}$.  We can write:
\begin{align}\label{NasSumOverRich}
\hat{N}_i| \{v_\nu\}_{\nu=1}^P \> = \sum_{\alpha} \frac{b^\dagger_i}{v_\alpha - \varepsilon_i} | \{ v_\nu \}_{\nu \neq \alpha} \>.
\end{align}
Considering that the operator $\hat{N}_i$ simply projects on the space of states that have $i$ occupied by a  Cooper pair and inspecting  Eq. (\ref{RichardsonState}), it is quite easy to understand how to derive (\ref{NasSumOverRich}). We shall not give a more explicit proof, but rather explain the different ingredients on the right hand side of. First note, that in order for the level $i$ to be occupied with a Cooper pair, one of the the operators $b_\nu$ in Eq. (\ref{RichardsonState}) must have hit the level $i$. The sum over $\alpha$ in (\ref{NasSumOverRich}) is a sum over all the possible such $\alpha$'s. The factor $\frac{1}{v_\alpha - \varepsilon_i}$ is inherited directly from $b^\dagger_\alpha$.  $b^\dagger_i$ in (\ref{NasSumOverRich}) is responsible for filling up the level $i$. All the rest of the levels have a chance to be filled by $b^\dagger_\nu$ except $\nu=\alpha$. This explains the state $ | \{ v_\nu \}_{\nu \neq \alpha} \>$ appearing in (\ref{NasSumOverRich}). This heuristic explanation may be translated directly into a rigorous proof, an alternative is to use the language of the algebraic Bethe ansatz to obtain the result, as done in [\onlinecite{Zhou:Links:McKenzie:Gould,Calabrese}].

We shall denote the state $b^\dagger_i  | \{ v_\nu \}_{\nu \neq \alpha} \>$ by $| \{ v_\nu \}_{\nu \neq \alpha} \cup \{\varepsilon_i\} \>$. We have:
\begin{align}\label{NasOverlap}
\hat{N}_i| \{v_\nu\}_{\nu=1}^P \> = \sum_{\alpha} \frac{1}{v_\alpha - \varepsilon_i} | \{ v_\nu \}_{\nu \neq \alpha} \cup \{\varepsilon_i\} \>
\end{align}
The state $| \{ v_\nu \}_{\nu \neq \alpha} \cup \{\varepsilon_i\} \> \equiv b^\dagger_i  | \{ v_\nu \}_{\nu \neq \alpha} \>$ can be thought as a Richardson state, with a set of rapidities, $v_\nu$, which do not satisfy Richardson's equations. This can be done  because of the following relation:
\begin{align}
| \{ v_\nu \}_{\nu \neq \alpha} \cup \{\varepsilon_i\} \> = \lim_{\epsilon\to 0} \epsilon | \{ v_\nu \}_{\nu \neq \alpha} \cup \{\varepsilon_i + \epsilon\} \> ,
\end{align}
where the state $| \{ v_\nu \}_{\nu \neq \alpha} \cup \{\varepsilon_i + \epsilon\} \> $ is given by (\ref{RichardsonState}).

Having written $\< \hat{N}_i \> $ as a sum over overlaps between Richardson states, we are ready to use Slavnov's formula to compute it. The results is:
\begin{align}
\< \hat{N}_i \> = \sum_\alpha \frac{\det A^{\left( \begin{smallmatrix} \alpha \\ i \end{smallmatrix} \right) }}{\det A},
\end{align}
where $A$ is given in (\ref{ADefinition}) and $A^{\left( \begin{smallmatrix} \alpha \\ i \end{smallmatrix} \right) }$ is $A$ with column $\alpha$ replaced by a column vector, $V^{(i)}$, this column vector being given by:
\begin{align}\label{Vdefinition}
V_\nu^{(i)} = \frac{1}{(v_\nu - \varepsilon_i)^2}.
\end{align}
More explicitly:
\begin{align}
A^{\left( \begin{smallmatrix} \alpha \\ i \end{smallmatrix} \right) }_{\mu,\nu} = \left\{ \begin{array}{lr}
V_\mu^{(i)} &  \nu= \alpha \\
A_{\mu,\nu} & \mbox{otherwise}
 \end{array}\right. .
\end{align}

By Cramer's rule the ratio of determinants can be computed as $\frac{\det A^{\left( \begin{smallmatrix} \alpha \\ i \end{smallmatrix} \right) }}{\det A} = \left( \left[ A^{-1}V^{(i)} \right]_\alpha \right)$. In order to be able to invert $A$, we note that $A$ has in fact a straightforward interpretation. Suppose that each one of the $v_\nu$'s are subjected to an external field $\delta h_\nu$. In order for them to remain in electrostatic equilibrium they must obey the equations :
\begin{align}\label{RichardsonWithExternal}
\delta \sum_{\mu\neq \nu} \frac{1}{ v_\nu-v_\mu}  - \frac{\delta}{2} \sum_{ i \in U } \frac{1}{v_\nu - \varepsilon_j }  - \frac{1}{g} = \delta h_\nu.
\end{align}
The shift of the $v_\nu$'s in linear order is a matrix multiplying the vector $\vec{\delta h}$ with components $\delta h_\nu$. This matrix turns out to be $\frac{2}{\delta} A^{-1}$. Indeed expanding (\ref{RichardsonWithExternal}) one obtains:
\begin{align}
\frac{\delta}{2} A_{i,j} \delta v_j = \delta h_i.
\end{align}
Suppose we shift  $\varepsilon_j$ by an amount $\delta \varepsilon_j$. This change can be represented as having $v_\nu$ experiencing an external field of $ \delta h_\nu = \frac{\delta}{2} \frac{\delta \varepsilon_j}{(v_\nu - \varepsilon_j)^2 } = \frac{\delta}{2} \delta \varepsilon_j  V^{(j)}_\nu$. Which implies:
\begin{align}\label{Amins1Asdvde}
\left[ A^{-1} V^{(i)} \right]_\mu = \frac{\partial v_\mu}{\partial \varepsilon_i },
\end{align}
which gives:
\begin{align}\label{NiAsSumOverDerivative}
\< \hat{N}_i \>  = \sum_\alpha \frac{\partial v_\alpha}{\partial \varepsilon_i}.
\end{align}
To compute $\frac{\partial v_\mu}{\partial \varepsilon_i}$, we resort to the results of section \ref{SpherAndEllipt}, where we have computed the coarse grained change in the potential due to a shift of a group of $\varepsilon$'s on the real axis. The jump discontinuity at $v_\mu$ of $\frac{\partial \phi(\xi)}{\partial \varepsilon}$ is equal to $2 \pi i \delta$ times an averaged $\frac{\partial v_\mu}{\partial \varepsilon_i}$ times the density of $v$'s. This is true if $v_\mu$ is on the arc or on the real axis, except that there is an extra contribution from unblocked $\varepsilon$. If we integrate over the jump discontinuity we obtain the sum on the right hand side of (\ref{NiAsSumOverDerivative}) with two caveats -- (1) the result is not a derivative with respect to a single $\varepsilon_i$ but a coarsed grained quantity and (2) the unblocked $\varepsilon$ add to the result. The two caveats amount to the fact that by integrating over the jump discontinuity of $\frac{\partial \phi(\xi)}{\partial \varepsilon}$ over the arcs and the real axis, we obtain $\hat{S}_z(\varepsilon)$ of definition (\ref{SzCoarseGrained}). Since the integral over the jump discontinuity is simply given by a contour integral surrounding both the real axis and the arcs, we have:
\begin{align}
\< \hat{S}_z(\varepsilon) \> = \frac{1}{2 \pi \delta i} \oint_\infty \frac{\partial \phi(\xi)}{\partial \varepsilon }  d\xi
\end{align}

In the spherical (one arc) case,  $\frac{\partial \phi(\xi)}{\partial \varepsilon }$ is given by (\ref{dphidepsiOneArc}), thus
\begin{align}
\<\hat{S}_z\>(\varepsilon)=\oint_\infty \frac{R_2(\xi)}{R_2(\varepsilon)(\xi-\varepsilon)}\frac{\rho(\varepsilon)}{2\pi}d\xi=\frac{\varepsilon-\mu}{R_2(\varepsilon)}{\rho(\varepsilon)},
\end{align}
which is the known BCS result, derived in this context by Richardson\cite{Richardson:Large:N}, based on more direct methods than the use of the Slavnov formula -- methods which are nevertheless harder to generalize to more complicated expectation values. We extend the result by considering two arced configurations. In this case,  $\frac{\partial \phi(\xi)}{\partial \varepsilon }$ is given by (\ref{dphidepsilon2Gap}), thus
\begin{align} \label{SzAsExplicitIntegral}
&\<\hat{S}_z(\varepsilon)\>=\frac{\rho(\varepsilon)}{R_4(\varepsilon)}\left\{\oint_{\infty}\frac{R_4(\xi)}{\xi-\varepsilon}d\xi-\oint_{u_{\infty}}\frac{ R_4(\xi) g(\xi)}{u'(\xi)}du(\xi) \right\}.
\end{align}
$u'(\xi)$ is computed making use of (\ref{xiofu}), to yield:
\begin{align}
u'(\xi)^{-1}=\wp(u(\xi)-u_\infty)-\wp(u(\xi)+u_\infty).
\end{align}
The first integral in (\ref{SzAsExplicitIntegral}) is to be taken over a large circle encompassing the arcs and $\varepsilon$ while the second integral is taken over the circle's image under $u(\xi)$. Performing the integration is a straightforward exercise in picking up the respective poles, the final result being:
\begin{align}
\< \hat{S}_z(\varepsilon) \> = \frac{(\varepsilon-\mu_1)(\varepsilon-\mu_2) + \frac{\Delta_1^2+\Delta_2^2}{2}   -2\wp(2u_\infty)-4\omega_1^{-1}\zeta(\frac{\omega_1}{2})  }{R_4(\varepsilon)}\rho(\varepsilon)
\end{align}

\subsection{Computation of $\< \hat{S}^\dagger(\varepsilon^*)\hat{S}(\varepsilon) \>$ \label{computeSdS}}
We now compute $\< \{v_\nu\}_{\nu=1}^P |\hat{S}^\dagger_{i^*} \hat{S}_i |\{v_\nu\}_{\nu=1}^P  \>$, assuming both $i$ and $i^*$ are unblocked levels.  Consider then
$\hat{S}^\dagger_{i^*} \hat{S}_i  |\{v_\nu\}_{\nu=1}^P  \> $.   $\hat{S}^\dagger_{i^*}$ projects  $|\{v_\nu\}_{\nu=1}^P  \>$ on the space in which $\varepsilon_{i^*}$ is empty and then fills it. The effect of $\hat{S}^\dagger_{i^*}$ can be simply achieved by adding $\varepsilon_{i^{*}}$ to the set of $v_\nu$, as this causes the level $i^*$ to be filled while ensuring that no $v_\nu$ hits $\varepsilon_{i^*}$.  Namely,
\begin{align}
\hat{S}_i \hat{S}^\dagger_{i^*}|\{v_\nu\}_{\nu=1}^P  \> = \hat{S}_i |\{v_\nu\}_{\nu=1}^P \cup \{\varepsilon_{i^*} \}  \>.
\end{align}
$\hat{S}_{i}$ projects  $|\{v_\nu\}_{\nu=1}^P  \cup \{\varepsilon_{i^*} \} \>$ on the space in which $\varepsilon_{i}$ is filled and then empties it. In formulas:
\begin{align}
\hat{S}_i |\{v_\nu\}_{\nu=1}^P \cup \{\varepsilon_{i^*} \}  \> = b_i \hat{N}_i |\{v_\nu\}_{\nu=1}^P \cup \{\varepsilon_{i^*} \}  \> = \sum_\alpha \frac{b_i}{v_\alpha - \varepsilon_i} |\{v_\nu\}_{\nu=1}^P \setminus \{v_\alpha\} \cup \{\varepsilon_{i^*} , \varepsilon_i \}  \>,
\end{align}
where in the last equality we have used the representation of  $\hat{N}_i$ as an overlap, Eq. (\ref{NasOverlap}). The following identity is easy to understand:
\begin{align}\label{Strick}
& {b_i} |\{v_\nu\}_{\nu=1}^P \setminus \{v_\alpha\} \cup \{\varepsilon_{i^*} , \varepsilon_i \}  \> = (1-\hat{N}_i) |\{v_\nu\}_{\nu=1}^P \setminus \{v_\alpha\} \cup \{\varepsilon_{i^*}  \}  \> =  \\ \nonumber
& = | \{v_\nu\}_{\nu=1}^P \setminus \{v_\alpha\} \cup \{\varepsilon_{i^*} \} \> - \sum_\beta \frac{1}{v_\beta - \varepsilon_i} | \{v_\nu\}_{\nu=1}^P \setminus \{v_\alpha , v_\beta\} \cup \{\varepsilon_{i^*} ,\varepsilon_i \} \>
\end{align}
making use again in the last equality of the representation of $\hat{N}_i$ as an overlap. We obtain:
\begin{align}\label{SdSAsOverlap}
& \< \{v_\nu\}_{\nu=1}^P  |\hat{S}_i \hat{S}^\dagger_{i^*}|\{v_\nu\}_{\nu=1}^P  \> = \\ \nonumber
& = \sum_\alpha \frac{1}{v_\alpha - \varepsilon_i} \left( \< \{v_\nu\}_{\nu=1}^P  | \{v_\nu\}_{\nu=1}^P \setminus \{v_\alpha\} \cup \{\varepsilon_{i^*} \} \> - \sum_\beta \frac{1}{v_\beta - \varepsilon_i} \< \{v_\nu\}_{\nu=1}^P  | \{v_\nu\}_{\nu=1}^P \setminus \{v_\alpha , v_\beta\} \cup \{\varepsilon_{i^*} ,\varepsilon_i \} \> \right).
\end{align}

Making Use Slavnov's formula, we are now ready to write the expectation value $\< \hat{S}^\dagger_{i^*} \hat{S}_i  \>$ as a determinant. The first term on the right hand side of (\ref{SdSAsOverlap}) goes along the same lines as the computation of $\<\hat{N}_i\>$, so we shall not repeat it here. The second term on the right hand side of (\ref{SdSAsOverlap}) has the added feature that it has two replacements $v_\alpha \to \varepsilon_{i^*}$ and $v_\beta \to \varepsilon_i$. This leads to the following equation:
\begin{align}\label{2by2Slavnov}
\< \{v_\nu\}_{\nu=1}^P  | \{v_\nu\}_{\nu=1}^P \setminus \{v_\alpha , v_\beta\} \cup \{\varepsilon_{i^*} ,\varepsilon_i \} \> = \frac{(v_\alpha - \varepsilon_i) (v_\alpha - \varepsilon_{i^*}) (v_\beta - \varepsilon_i) (v_\beta - \varepsilon_{i^*}) }{(v_\alpha-v_\beta)(\varepsilon_i - \varepsilon_{i^*})}{\det A^{\left( \begin{smallmatrix} \alpha &  \beta \\ i^* & i \end{smallmatrix} \right)} } ,
\end{align}
where
\begin{align}
A^{\left( \begin{smallmatrix} \alpha &  \beta \\ i^* & i \end{smallmatrix} \right)}_{\mu,\nu} = \left\{ \begin{array}{lr}
V^{(i^*)}_\mu & \nu = \alpha \\
V^{(i)}_\mu & \nu = \beta \\
A_{\mu, \nu} & \mbox{otherwise}
\end{array} \right.
\end{align}
Cramer's rule for $A^{\left( \begin{smallmatrix} \alpha &  \beta \\ i^* & i \end{smallmatrix} \right)}$ reads
\begin{align}
\frac{\det A^{\left( \begin{smallmatrix} \alpha &  \beta \\ i^* & i \end{smallmatrix} \right)}}{\det A} = \det \left( \begin{array}{cc}
\left( A^{-1} V^{(i^*)} \right)_\alpha &  \left( A^{-1} V^{(i^*)} \right)_\beta \\
\left( A^{-1} V^{(i)} \right)_\alpha &  \left( A^{-1} V^{(i)} \right)_\beta \end{array}\right),
\end{align}
which according to (\ref{Amins1Asdvde}) reads:
\begin{align} \label{2by2AasDetdVdEpsilon}
\frac{\det A^{\left( \begin{smallmatrix} \alpha &  \beta \\ i^* & i \end{smallmatrix} \right)}}{\det A} = \det \left( \begin{array}{cc}
\frac{\partial v_\alpha}{\partial \varepsilon_{i^*}} &  \frac{\partial v_\beta}{\partial \varepsilon_{i^*}} \\
\frac{\partial v_\alpha}{\partial \varepsilon_{i}} &   \frac{\partial v_\beta}{\partial \varepsilon_{i}} \end{array} \right)
\end{align}
Combining (\ref{SdSAsOverlap}), (\ref{2by2Slavnov}) and (\ref{2by2AasDetdVdEpsilon}) we obtain:
\begin{align}\label{SdaggerSDiscrete}
\< \hat{S}^\dagger_{i^*} \hat{S}_i \> = \sum_{ \alpha } \frac{v_\alpha - \varepsilon_{i^*} }{ v_\alpha - \varepsilon_i } \frac{\partial v_\alpha}{\partial \varepsilon_{i^*}} - \sum_{ \alpha, \beta }  \frac{(v_\alpha -\varepsilon_{i^*})(v_\beta -\varepsilon_{i^*})}{(v_\alpha - v_\beta)(\varepsilon_i -\varepsilon_{i^*})} \left( \frac{\partial v_\alpha}{ \partial \varepsilon_{i^*} }  \frac{\partial v_\beta}{ \partial \varepsilon_{i} }  - \frac{\partial v_\alpha}{ \partial \varepsilon_{i} }  \frac{\partial v_\beta}{ \partial \varepsilon_{i^*} } \right).
\end{align}
We  now  need to take the continuum limit of the expression. This is achieved by coarse graining the quantities $\hat{S}_i$ and $\hat{S}^\dagger_{i^*}$. Explicitly, the coarse graining reads $\hat{S}^\dagger(\varepsilon^*) = \frac{1}{2A}\sum_{|\varepsilon_{i^*} - \varepsilon^*| < A \delta} \hat{S}^\dagger_{i^*}$ and $\hat{S}(\varepsilon) = \frac{1}{2A}\sum_{|\varepsilon_i - \varepsilon^*| < A \delta} \hat{S}_{i}$. The first sum in (\ref{SdaggerSDiscrete}) has the following continuum limit:
\begin{align}
\sum_{\alpha } \frac{v_\alpha - \varepsilon_{i^*} }{ v_\alpha - \varepsilon_i } \frac{\partial v_\alpha}{\partial \varepsilon^*} \longrightarrow \rho_U (\varepsilon^*) \oint_\Gamma \frac{(\xi^*-\varepsilon^*)}{(\xi^* - \varepsilon)} \frac{\partial \phi(\xi^*)}{\partial \varepsilon^*}  \frac{d\xi^*}{2\pi i \delta},
\end{align}
where the integral encircles the arcs, but no part of the real axis (except the intersection of the arc with the real axis). $\rho_U(\varepsilon^*)$ denotes the average occupation of unblocked levels at $\varepsilon^*$. This term appears because the unblocked $i$'s (and only them) must be summed over in the  coarse graining procedure. Indeed $i$ and $i^*$ are {\it assumed} to be unblocked in (\ref{SdaggerSDiscrete}) and if either one is blocked the correlation function is obviously zero. Another point to note is that since the integral is taken over $\Gamma$ the contribution of $\frac{\partial v_\alpha}{\partial \varepsilon^*}$ is neglected for real $v_\alpha$. However this poses no difficulty, since only real $v_\alpha$ near $\varepsilon_{i^*}$ are affected by a change of $\varepsilon_{i^*}$, and  their contribution to the sum is suppressed by a factor $(v_\alpha - \varepsilon_{i^*})$. Namely, this contribution does not survive in the continuum limit.

Treating now the continuum limit of the second, double, sum in (\ref{SdaggerSDiscrete}) we obtain:
\begin{align}
&\sum_{ \alpha, \beta }  \frac{(v_\alpha -\varepsilon_{i^*})(v_\beta -\varepsilon_{i^*})}{(v_\alpha - v_\beta)(\varepsilon_i -\varepsilon_{i^*})} \left( \frac{\partial v_\alpha}{ \partial \varepsilon_{i^*} }  \frac{\partial v_\beta}{ \partial \varepsilon_{i} }  - \frac{\partial v_\alpha}{ \partial \varepsilon_{i} }  \frac{\partial v_\beta}{ \partial \varepsilon_{i^*} } \right) \longrightarrow \nonumber \\
& - 2 \rho_v (\varepsilon) \oint_\Gamma \frac{(\xi^*-\varepsilon^*)}{(\xi^* - \varepsilon)} \frac{\partial \phi(\xi^*)}{\partial \varepsilon^*}  \frac{d\xi^*}{2\pi i \delta} + \oint_\Gamma \oint_\Gamma\frac{(\xi - \varepsilon^*)(\xi^* - \varepsilon^*)}{(\xi-\xi^*)(\varepsilon - \varepsilon^*)}  \left( \frac{\partial \phi(\xi) }{\partial \varepsilon} \frac{\partial \phi(\xi^*)}{\partial \varepsilon^*} -\frac{\partial \phi(\xi^*) }{\partial \varepsilon} \frac{\partial \phi(\xi)}{\partial \varepsilon^*}  \right)\frac{d\xi}{2\pi i \delta} \frac{d\xi^*}{2\pi i \delta}.
\end{align}
The double integral on the right hand side is the obvious continuum limit of the left hand side. The single integral takes into account the contribution of $v$'s near $\varepsilon_i$, which is neglected in the double integral, which is performed over $\Gamma$.  This contribution is naturally proportional to $\rho_v(\varepsilon)$, the average occupation of $v$'s near $\varepsilon$ (again being related to the fact that the $v$'s move rigidly with the $\varepsilon$'s).  The contribution of $v$'s near $\varepsilon_{i^*}$ is suppressed by the factor $(v_\alpha - \varepsilon_{i^*})(v_\beta - \varepsilon_{i^*})$ and thus need not be taken.

Since the average (coarse-grained) charge, $\rho$ is defined as $\rho = \frac{\rho_U - 2 \rho_v}{2}$ we obtain an expression in the continuum limit in the following form:
\begin{align}
\<\hat{S}^\dagger(\varepsilon) \hat{S}(\varepsilon) \> =2 \rho(\varepsilon) \oint_\Gamma \frac{(\xi^*-\varepsilon^*)}{(\xi^* - \varepsilon)} \frac{\partial \phi(\xi^*)}{\partial \varepsilon^*}  \frac{d\xi^*}{2\pi i \delta} - \oint_\Gamma \oint_\Gamma\frac{(\xi - \varepsilon^*)(\xi^* - \varepsilon^*)}{(\xi-\xi^*)(\varepsilon - \varepsilon^*)}  \left( \frac{\partial \phi(\xi) }{\partial \varepsilon} \frac{\partial \phi(\xi^*)}{\partial \varepsilon^*} -\frac{\partial \phi(\xi^*) }{\partial \varepsilon} \frac{\partial \phi(\xi)}{\partial \varepsilon^*}  \right)\frac{d\xi}{2\pi i \delta} \frac{d\xi^*}{2\pi i \delta}
\end{align}
We shall not perform the integrals explicitly, since the result is not very illuminating. We shall proceed instead to giving the general expression for the expectation value of any number of operators.

\section{Computation of Expectation values: General formula}
We now compute a general expectation value featuring any fixed (not scaling with $\frac{1}{\delta}$) number of operators. The first thing to do is to write such an expectation value as an overlap of Richardson states. This is done either by invoking concepts related to the algebraic Bethe ansatz, or using repeatedly the tricks of subsections \ref{ComputeSz} and \ref{computeSdS}. The result is:
\begin{align}\label{generalExpectAsOverlap}
&\< \hat{S}_{i_1}  \hat{S}_{i_2} \dots \hat{S}_{i_n} \hat{S}^\dagger_{i^*_1}  \dots \hat{S}^\dagger_{i^*_n}  \hat{N}_{j_1} \dots \hat{N}_{j_m}\> = \\ \nonumber & = \sum_{k=1}^n \,\, \sum_{m_1<m_2<\dots<m_k} \,\, \sum_{\nu_1,\dots,\nu_{n+m+k}}\frac{ (-)^k }{\prod_{l=1}^n (v_{\nu_l} - \varepsilon_{i_l}) \prod_{l=1}^k (v_{\nu_{n+l}} - \varepsilon_{i_{m_l}})\prod_{i=l}^m(v_{\nu_{n+k+l}}-\varepsilon_{j_l})} \times \\
& \times \frac{\left\< \{v_\mu\}_{\mu=1}^P\right|\left. \{v_\mu\}_{\mu=1}^P \setminus \{v_{\nu_l}\}_{l=1}^{n+m+k} \cup \{ \varepsilon_{i^*_l}\}_{l=1}^n \cup \{ \varepsilon_{i_{m_l}}\}_{l=1}^k \cup \{ \varepsilon_{j_l}\}_{l=1}^m \right\>}{\left\< \{v_\mu\}_{\mu=1}^P\right|\left. \{v_\mu\}_{\mu=1}^P \right\>} \nonumber .
\end{align}
The factor $(-)^k$ comes from a straightforward inclusion-exclusion principle, or alternatively from expanding the product  $\prod_{j=1}^n(1-\hat{N}_{i_j})$, whose origin is the same as the origin of $(1-\hat{N}_i)$ appearing in (\ref{Strick}).

The overlaps appearing in (\ref{generalExpectAsOverlap}) can be easily computed using Slavnov's formula, with the result:
\begin{align}
\frac{\left\< \{v_\mu\}_{\mu=1}^P\right|\left. \{v_\mu\}_{\mu=1}^P \setminus \{v_{\nu_l}\}_{l=1}^{s}  \cup \{ \varepsilon_{k_l}\}_{l=1}^s     \right\>}{\left\< \{v_i\}_{i=1}^P\right|\left. \{v_i\}_{i=1}^P \right\>} = \frac{\prod_{l,m}(v_{\nu_l}-\varepsilon_{k_m}) }{\prod_{m<n}(\varepsilon_{k_m} -\varepsilon_{k_n}) \prod_{m<n} (v_{\nu_m}-v_{\nu_n})}\frac{\det A^{\left( \begin{smallmatrix} \nu_1 & \nu_2 & \dots  & \nu_s \\ k_1 & k_2 & \dots & k_s \end{smallmatrix} \right)}}{\det A},
\end{align}
in which $A^{\left( \begin{smallmatrix} \nu_1 & \nu_2 & \dots  & \nu_s \\ k_1 & k_2 & \dots & k_s \end{smallmatrix} \right)}$ is defined as:
\begin{align}\label{AalphaDefinition}
 A^{\left( \begin{smallmatrix} \nu_1 & \nu_2 & \dots  & \nu_s \\ k_1 & k_2 & \dots & k_s \end{smallmatrix} \right)}_{\gamma,\delta} = \left\{ \begin{array}{lr}
V^{(k_i)}_\gamma & \exists  i, \delta = \nu_i \\
A_{\gamma, \delta} & \mbox{otherwise}
\end{array} \right. ,
\end{align}
with
\begin{align} \label{VDefinition}
V^{(k_i)}_\gamma = \frac{1}{(v_\gamma-\varepsilon_{k_i})^2} .
\end{align}
$A^{-1}$  is given an electrostatic interpretation just as above to yield:
\begin{align} \label{detAisDETdEdepsilon}
\frac{\det A^{\left( \begin{smallmatrix} \nu_1 & \nu_2 & \dots  & \nu_s \\ k_1 & k_2 & \dots & k_s \end{smallmatrix} \right)}}{\det A} = \det_{i,j} \left(\left[ A^{-1} V^{(k_i)}\right]_{\nu_j} \right) = \det_{i,j} \frac{\partial v_{\nu_j}}{\partial \varepsilon_{k_i}}.
\end{align}
And the whole expectation value has the following continuum limit version:
\begin{align}\label{generalExpectationContinuum}
&\<\hat{S}(\varepsilon_{k_1})  \hat{S}(\varepsilon_{k_2}) \dots \hat{S}(\varepsilon_{k_n}) \hat{S}^\dagger(\varepsilon_{k_{n+1}})\dots \hat{S}^\dagger(\varepsilon_{k_{2n}}) \hat{S}_z(\varepsilon_{k_{2n+1}}) \dots \hat{S}_z(\varepsilon_{k_{2n+m}})\> = \\
& = \sum_{k=0}^n (-)^k \sum_{I \in P_k } \left[ \prod_{i \in I} \oint_\Gamma \frac{d\xi_{i}}{2\pi} \right] \prod_{\begin{array}{c} (i, j)\in I^2 \\ i < j \end{array}}\frac{(\xi_{i} - \varepsilon_{k_{j}})(\xi_{j} - \varepsilon_{k_{i}})}{(\xi_{i} - \xi_{j}) (\varepsilon_{k_i} - \varepsilon_{k_j} )} \prod_{i=n+1 }^{2n} \frac{\xi_{i} - \varepsilon_{k_{i}}}{\xi_{i} - \varepsilon_{k_{i-n}}}\det_{ (i,j) \in I^2} \frac{\delta \phi(\xi_{j})}{\delta  \varepsilon_{k_j}}, \nonumber
\end{align}
where
\begin{align}
P_k = \left\{ I \subseteq \{1,2,\dots,2n+m \}|  I = \left\{j_1,j_2,\dots,j_k,n+1,n+2,\dots,2n+m\right\},\mbox{ where }  1\leq j_1<j_2<\dots<j_k\leq n \right\}.
\end{align}

The difficult object to compute in (\ref{generalExpectationContinuum}) is
$\det_{ i,j} \frac{\delta \phi(\xi_{j})}{\delta  \varepsilon_{k_j}}$.
Eq. (\ref{dphidepsilon2Gap}) shows that, up to a multiplication of rows an columns by constant factors, the matrix
$\frac{\delta \phi(\xi_{j})}{\delta  \varepsilon_{k_j}}$
 has a  part which is a Cauchy matrix, the Cauchy matrix being given by:
\begin{align}
C_{i,j} & = \frac{1}{\xi_{j} - \varepsilon_{k_i}}
\end{align}
Indeed
\begin{align}\label{detphiasCplusG}
\det_{(i,j)\in I^2} \frac{\delta \phi(\xi_{j})}{\delta  \varepsilon_{k_j}} =\prod_{i \in I} \frac{\rho(\varepsilon_{k_i}) R_4(\xi_{i}) }{R_4(\varepsilon_{k_i})} \det_{(i,j)\in I^2} \left[C + G\right]_{i,j},
\end{align}
where
\begin{align}
G_{i,j} &= g(\xi_{j}).
\end{align}
We use this fact to write:
\begin{align}\label{detphi}
\det_{(i,j)\in I^2} \frac{\delta \phi(\xi_{j})}{\delta  \varepsilon_{k_j}} = \prod_{i \in I } \frac{\rho(\varepsilon_{k_i}) R_4(\xi_{i}) }{R_4(\varepsilon_{k_i})}  \det_{(i,j)\in I^2} (C_{i,j}) \det_{(i,j)\in I^2} \left( \delta_{i,j} - \sum_{l \in I}  C_{i,l}^{-1} G_{l,j}   \right)
\end{align}
The determinant of the Cauchy matrix is known to be given by :
\begin{align}\label{CauchyDet}
\det_{(i,j)\in I^2} (C_{i,j}) = \frac{ \prod_{(i,j) \in I^2,  i < j} \left[ ( \varepsilon_{k_i} - \varepsilon_{k_j}) (\xi_i -
\xi_j) \right]}{\prod_{(i,j) \in I^2} (\xi_i  - \varepsilon_{k_j})  }
\end{align}
The inverse of the Cauchy matrix is also known, this is given by:
\begin{align}
C^{-1}_{i,j} = \frac{\prod_{l \in I} \left[ (\xi_{i} - \varepsilon_{k_l})(\xi_{l} - \varepsilon_{k_j})\right]}{\prod_{l \in I, l \neq i}(\xi_{i}-\xi_{l}) \prod_{l \in I, l \neq j}(\varepsilon_{k_j} - \varepsilon_{k_l} )}\frac{1}{\xi_{i} - \varepsilon_{k_j}}.
\end{align}
Note however, that this is the inverse of $C$ when it is understood that the indices run only over the set $I$, namely:
\begin{align}
\forall (i,j)\in I^2, \quad \sum_{l\in I} C^{-1}_{i,l}C_{l,j} = \delta_{i,j}.
\end{align}
Making use of an algebraic identity:
\begin{align}
\sum_{j\in I} \frac{\prod_{l\in I} (\xi_{l} - \varepsilon_{k_j})}{ \prod_{l\in I, l \neq  j}(\varepsilon_{k_j} - \varepsilon_{k_l} )}\frac{1}{\xi_{i} - \varepsilon_{k_j}} =  \oint \frac{\prod_{l\in I , l \neq i} (x- \xi_{l} )}{ \prod_{l\in I}(  x -\varepsilon_{k_l} )} \frac{dx}{2\pi i} =  1,
\end{align}
one obtains that the matrix $\left[ C^{-1}G\right]_{i,j}$ is diadic:
\begin{align}
\left[ C^{-1}G\right]_{i,j} = \frac{\prod_{l \in I}(\xi_{i} - \varepsilon_{k_l})}{\prod_{l\in I, l\neq i}(\xi_{i}-\xi_{l})} g(\xi_{j}).
\end{align}
If $r$ and $s$ are column vectors and $F$ is a diadic matrix formed out of them, $F=r s^t$, then $\det(\mathds{1} + F) = 1+ r^t s$. This allows us to write:
\begin{align}\label{sumgascontourwith1}
\det_{(i,j) \in I^2} \left( \delta_{i,j} -  \sum_{l\in I} C^{-1}_{i,l} G_{l,j}   \right) = 1 - \sum_i  \frac{\prod_{l\in I}(\xi_{i} - \varepsilon_{k_l})}{\prod_{l\in I, l\neq i}(\xi_{i}-\xi_{l})} g(\xi_{i}) = 1 - \oint_\mathcal{C}  \frac{\prod_{l\in I}(x - \varepsilon_{k_l})}{\prod_{l\in I}(x-\xi_{l})} g(x) \frac{dx}{2\pi i},
\end{align}
where the contour, $\mathcal{C}$,  on the right hand side encircles all $\xi_i$'s. We let this contour be composed of two parts, a large circle traversed counterclockwise around infinity, and a contour, $\Gamma$, traversed clockwise around the arcs, $\oint_\mathcal{C} = \oint_\infty - \oint_\Gamma$. The integral over the large circle can be done immediately by expanding its radius to infinity. This integral can be easily seen to be equal to $1$. We are thus left only with the integral over the contour $\Gamma$:
\begin{align}\label{sumgascontour}
\det_{(i,j) \in I^2} \left( \delta_{i,j} -  \sum_{l\in I} C^{-1}_{i,l} G_{l,j}   \right) =   \oint_\Gamma  \frac{\prod_{l\in I}(x - \varepsilon_{k_l})}{\prod_{l\in I}(x-\xi_{l})} g(x) \frac{dx}{2\pi i},
\end{align}
where the integral is taken counterclockwise around the arcs. Combining (\ref{generalExpectationContinuum}), (\ref{detphi}), (\ref{CauchyDet})  and (\ref{sumgascontour}) we obtain:
\begin{align}\label{AlmostFinalResult}
&\<\hat{S}(\varepsilon_{k_1})  \hat{S}(\varepsilon_{k_2}) \dots \hat{S}(\varepsilon_{k_n}) \hat{S}^\dagger(\varepsilon_{k_{n+1}})\dots \hat{S}^\dagger(\varepsilon_{k_{2n}}) \hat{S}_z(\varepsilon_{k_{2n+1}}) \dots \hat{S}_z(\varepsilon_{k_{2n+m}})\>  =  \\ \nonumber
& \left[ \prod_{i=1}^{2n+m}  \rho (\varepsilon_i)\right]  \oint_\Gamma g(x)  \left[ \prod_{i=1}^n \left( 2 - \oint_\Gamma \frac{ R_4(\xi_i) (x-\varepsilon_i)  }{R_4(\varepsilon_i)(x-\xi_i)(\xi_i - \varepsilon_i)} \frac{d\xi_i}{2 \pi i \delta} \right)\prod_{i=n+1}^{2 n} \left( \oint_\Gamma \frac{ R_4(\xi_i) (x-\varepsilon_i)  }{R_4(\varepsilon_i)(x-\xi_i)(\xi_i - \varepsilon_{i-n})} \frac{d\xi_i}{2 \pi i \delta}  \right) \right. \times  \\ \nonumber
& \times  \left. \prod_{i=2n+1}^{2 n+m} \left( \oint_\infty \frac{ R_4(\xi_i) (x-\varepsilon_i)  }{R_4(\varepsilon_i)(x-\xi_i)(\xi_i - \varepsilon_{i})} \frac{d\xi_i}{2 \pi i \delta} \right) \right] \frac{dx}{2 \pi i}
\end{align}
The result has a convenient almost-factorized form. Performing the integrals over $\xi$'s directly we obtains our main result:
\begin{align}\label{FinalResult}
&\<\hat{S}(\varepsilon_{k_1})  \hat{S}(\varepsilon_{k_2}) \dots \hat{S}(\varepsilon_{k_n}) \hat{S}^\dagger(\varepsilon_{k_{n+1}})\dots \hat{S}^\dagger(\varepsilon_{k_{2n}}) \hat{S}_z(\varepsilon_{k_{2n+1}}) \dots \hat{S}_z(\varepsilon_{k_{2n+m}})\>  =  \\ \nonumber
& \left[ \prod_{i=1}^{2n+m}  \rho (\varepsilon_i) \right]  \oint_\Gamma g(x)  \left[ \prod_{i=1}^n \left( 1 - \left( \frac{\ (\mu_1+\mu_2 -(x+\varepsilon_i)) (x-\varepsilon_i)  }{ R_4(\varepsilon_i)}   \right)^2 \right)  \frac{R_4(\varepsilon_{i})}{x-\varepsilon_{i}} \times \right.\\ \nonumber
& \left. \times \prod_{i=n+1}^{2 n} \frac{x-\varepsilon_i }{ R_4(\varepsilon_i)}   \prod_{i=2n+1}^{2 n+m} \left( \frac{ (\mu_1+\mu_2 -(x+\varepsilon_i)) (x-\varepsilon_i)  }{R_4(\varepsilon_i)} \right) \right] \frac{dx}{2 \pi i },
\end{align}
where $g(x)$ is given by (\ref{gDefinition}) and $R_4(\xi)$ is given by (\ref{Rdefinition}). The result appears formidable, but consists  only of a single  contour integral over known functions. The computation of this integral involves finding residues of the integrand, which is a mechanical task, easily performed by mathematical software, such that more explicit expressions can be derived for given $n$ and $m$.

In case one arc vanishes, e.g.,  $\Delta_2\to0$,  the function $g(\xi)$ can be shown to take the limit $g(\xi) \to \frac{1}{\xi - \mu_2}$. In this case the  integral over $x$ in (\ref{AlmostFinalResult}) can be taken by shrinking the contour of integration to a point, $\mu_2$. This amounts to a substitution $x\to \mu_2$, and gives the BCS result:
\begin{align}\label{BCSResult}
&\<\hat{S}(\varepsilon_{k_1})  \hat{S}(\varepsilon_{k_2}) \dots \hat{S}(\varepsilon_{k_n}) \hat{S}^\dagger(\varepsilon_{k_{n+1}})\dots \hat{S}^\dagger(\varepsilon_{k_{2n}}) \hat{S}_z(\varepsilon_{k_{2n+1}}) \dots \hat{S}_z(\varepsilon_{k_{2n+m}})\>  =  \\ \nonumber
& =      \prod_{i=1}^{2n} \frac{\Delta   }{ R_2(\varepsilon_i)} \rho (\varepsilon_i) \prod_{i=2n}^{2n+m} \frac{(\varepsilon_i - \mu ) }{ R_2(\varepsilon_i)} \rho (\varepsilon_i) ,
\end{align}
Note however that this way to obtain the BCS result is not general. The point $\mu_2$ on the real axis is special, and has the property that the far-field ($h(\mu_2 +i 0^+) + h(\mu_2 -i 0^+) =0 $) vanishes at this point. Not all solutions with one arc obey this constraint. The general way to obtain the BCS result is to rather take  expression (\ref{dphidepsiOneArc}) for $\frac{\partial \phi(\xi_i)}{\partial \varepsilon_j}$ as a starting point. This amounts to taking $G=0$ in (\ref{detphiasCplusG}). It is easy then to proceed since the integral over $g(x)$ does not show up and in fact the one-arc version of (\ref{AlmostFinalResult}) takes the simplified form:
\begin{align}\label{AlmostFinalResult1Arc}
&\<\hat{S}(\varepsilon_{k_1})  \hat{S}(\varepsilon_{k_2}) \dots \hat{S}(\varepsilon_{k_n}) \hat{S}^\dagger(\varepsilon_{k_{n+1}})\dots \hat{S}^\dagger(\varepsilon_{k_{2n}}) \hat{S}_z(\varepsilon_{k_{2n+1}}) \dots \hat{S}_z(\varepsilon_{k_{2n+m}})\>  =  \\ \nonumber
& \prod_{i=1}^{2n+m}  \rho (\varepsilon_i)  \prod_{i=1}^n \left( 2 - \oint_\Gamma \frac{ R_2(\xi_i) }{R_2(\varepsilon_i)(\xi_i - \varepsilon_i)} \frac{d\xi_i}{2 \pi i \delta} \right)\prod_{i=n+1}^{2 n} \left( \oint_\Gamma \frac{ R_2(\xi_i)  }{R_2(\varepsilon_i)(\xi_i - \varepsilon_{i-n})} \frac{d\xi_i}{2 \pi i \delta}  \right)   \prod_{i=2n+1}^{2 n+m} \left( \oint_\infty \frac{ R_2(\xi_i)  }{R_2(\varepsilon_i)(\xi_i - \varepsilon_{i})} \frac{d\xi_i}{2 \pi i \delta} \right) .
\end{align}
The integrals can be explicitly taken to give (\ref{BCSResult}).

\section{Conclusion}
We have shown how to compute correlation functions in the thermodynamic limit of the Richardson model. We gave explicit results for the case of one arc and two arcs. The one arc results converge with the BCS result, as expected. The correlation functions factorize into independent factors corresponding to each one of the operators in the correlation function. In the two arc case the factorization property disappears. Instead, the result, Eq. (\ref{FinalResult}), is given as a contour integral over an auxiliary variable $x$, which has a factorized form, where again each factor corresponds to an operator in the correlation function.

It is interesting to see whether our results may also be obtained from a semi-classical approach, following the works of [\onlinecite{kuznetsov, Kuznetsov2}]. In this approach the semiclassical analogues  \cite{Spivak:Levitov:Barankov}, $S^+$, $S^-$, $S_z$, of the operators $\hat{S}^\dagger$, $\hat{S}$, $\hat{S}_z$, respectively, are considered. These are shown to obey a classical integrable nonlinear equation. Being integrable, the equation may be solved\cite{kuznetsov, Kuznetsov2, Spivak:Levitov:Barankov}. It may then be possible to compute correlation functions in a semiclassical approach. In the case where the order parameter $\Delta(t)$ is time independent, this approach converges with the BCS approach, producing correct results. In case the order parameter $\Delta(t)$ is time dependent, it may be more delicate to justify the semiclassical approach for the computation of expectation values. The question of validity notwithstanding, our final result is suggestive of such an approach. Indeed, it may be that by a change of variables the integral over $x$ turns into an integral over time, the periodicity of the integration contour over $x$ related to the periodicity of a semiclassical solution, in which case our result may turn simply into a time average of the product of the respective semiclassical spin components, $S^+$, $S^-$, $S_z$.

A more challenging task, one we intend to pursue in future studies, is the calculation of matrix elements. For example $\<v| c^\dagger_{j,\sigma} c_{j,\sigma} |w\>$ or $\<v|c^\dagger_{j,+} c^\dagger_{j,-} |w\>$, between two different eigen-states, $|v\>$, and $|w\>$. These matrix elements are important in predicting the dynamics of Richardson's state and in revealing its quantum coherence properties in different physical situations. Indeed $\<v| c^\dagger_{j,\sigma} c_{j,\sigma} |w\>$ is related to the transition rate between state $|v\>$ and $|w\>$ under a perturbation $c^\dagger_{j,\sigma} c_{j,\sigma}$. Such objects appear in the computation of the Fermi golden rule rate due to, e.g., phonon scattering. Note that The object $\<v| c^\dagger_{j,\sigma} c_{j,\sigma} |w\>$, doesn't have a natural semiclassical counterpart, as the operators $c^\dagger_{j,\sigma}$, $c_{j,\sigma}$, are not simply related to $S^+$, $S^-$, $S_z$.  The other matrix element mentioned, namely $\<v|c^\dagger_{j,+} c^\dagger_{j,-} |w\>$, appears naturally when one attempts to compute the tunneling of pairs into a superconductor in state $|v\>$. Such a computation appears in treating the Josephson effect or Andreev reflection. A Josephson effect setup is the obvious choice to measure the time dependent order parameter $\Delta(t)$ in an experiment.

The research was supported by a grant from the Israel Science Foundation, grant no. 852/11. Additional support was provided by a grant from the Binational Science Foundation, grant no. 2010345. We acknowledge discussions with A. Nahum, B. Spivak, O. Agam and D. Orgad.

\bibliographystyle{apsrev4-1}
\bibliography{mybib}

\end{document}